\documentclass[prb,aps,twocolumn,amsmath,amssymb,notitlepage]{revtex4-1}

\usepackage[hidelinks]{hyperref}

\usepackage{mathtools}
\usepackage{amssymb}
\usepackage{amsfonts}
\usepackage{amsmath}
\usepackage{amsthm}
\usepackage{latexsym}
\usepackage{dcolumn}

\usepackage{amsmath}
\usepackage{braket}
\usepackage{graphicx}
\usepackage{framed}
\usepackage{wrapfig}
\usepackage[table]{xcolor}

\usepackage{psfrag,pstricks}

\definecolor{shadecolor}{RGB}{242,244,242}
\definecolor{headerrow}{RGB}{230,230,230}
\definecolor{row1}{RGB}{250,250,255}
\definecolor{row2}{RGB}{245,245,255}
\rowcolors{1}{row1}{row2}

\newcommand*{\etc}[1]{\textcolor{black}{#1}}

\newcounter{boxcount}
\newcommand{\boxnum}{\refstepcounter{boxcount}\theboxcount~}

\begin{document}

\title{Roads towards fault-tolerant universal quantum computation}

\author{Earl T. Campbell}
\affiliation{Department of Physics and Astronomy, University of Sheffield,
Sheffield, UK}
\author{Barbara M. Terhal}
\affiliation{JARA Institute for Quantum Information, RWTH Aachen University, 52056 Aachen, Germany}
\author{Christophe Vuillot}
\affiliation{JARA Institute for Quantum Information, RWTH Aachen University, 52056 Aachen, Germany}
\date{\today}

\begin{abstract}
Current experiments are taking the first steps toward noise-resilient logical qubits.
Crucially, a quantum computer must not merely store information, but also process it.
A fault-tolerant computational procedure ensures that errors do not multiply and spread. This review compares the leading proposals for promoting a quantum memory to a quantum processor. We compare magic state distillation, color code techniques and other alternative  ideas, paying attention to relative resource demands. We discuss the several no-go results which hold for low-dimensional topological codes and outline the potential rewards of using high-dimensional quantum (LDPC) codes in modular architectures.
\end{abstract}


\maketitle

 
 


\section{Introduction}
\label{sec:intro} 




The next decade will likely herald controllable quantum systems with 30 or more physical qubits on various quantum technology platforms, such as ion-traps \cite{HRB:ions,ballance:highFidelity} or superconducting qubits \cite{DS:outlook}.
It may be difficult to simulate such partially-coherent, dynamically-driven, many-body systems on a classical computer, since the elementary two-qubit entangling gate time can be as much as a 1000 times faster than the single-qubit dephasing and relaxation times ($T_2$ and $T_1$).
On the other hand, a system in which one out of a 1000 components fails is unlikely to perform well in executing large quantum algorithms designed for fault-free components.
We must either figure out what computational tasks a noisy many-body quantum system can perform well or we use partially-coherent qubits as the elementary constituents of more robust logical qubits through quantum error correction.
The choice of quantum error correcting architecture determines all operations at the physical hardware level.
It constrains the compilation from quantum software and quantum algorithms to actions on elementary qubits in hardware. For superconducting qubits, efforts to build a first logical qubit of the surface code are underway at places such as IBM Research \cite{corcoles:surface, IBM:surface}, UCSB in partnership with Google \cite{kelly:repetition} and the TU Delft \cite{riste+:3bit}. 


\begin{figure*}[thb]
\centering
\includegraphics{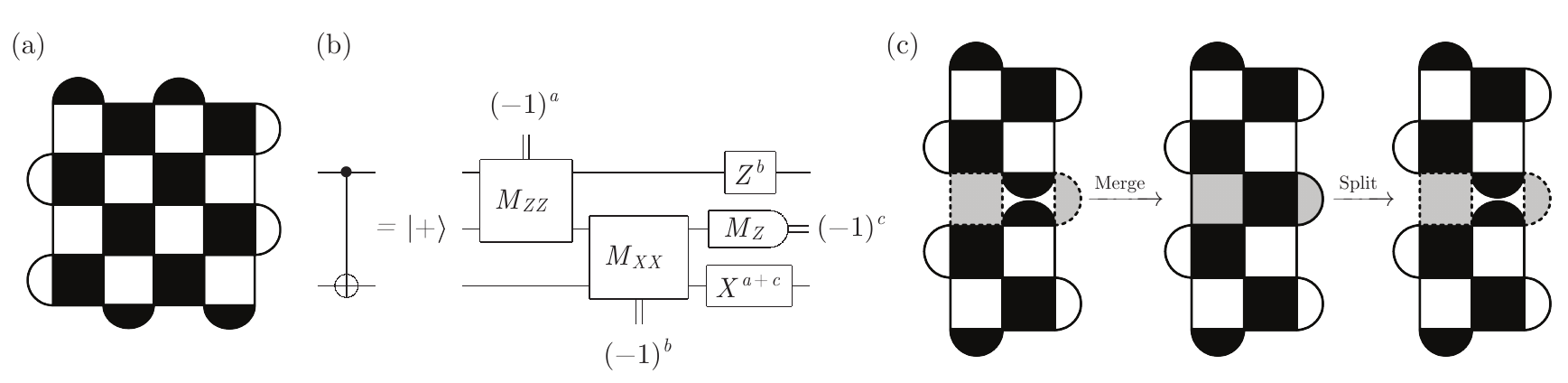}
\caption{(a) One logical qubit is encoded in a surface code sheet consisting of $d^2$ physical qubits at the vertices of the lattice ($d=5$ in the Figure). Black faces represent $X$-parity checks and white faces represent $Z$-parity checks on the qubits. (b) A CNOT circuit is equivalent to performing non-destructive $XX$ and $ZZ$ measurements for the control and target qubits with an ancillary qubit.  Lattice code surgery provides a method for non-destructively measuring a logical $ZZ$ or $XX$ between two encoded qubits. (c) Lattice code surgery between two surface code sheets which realizes a logical $ZZ$ measurement. By measuring and merging the parity checks between two surface code sheets, we merge the two sheets into one. At the same time one learns the value of the logical $ZZ$ as it equals the product of the newly measured grey faces. The two sheets can then be split again for further operations.}
\label{fig:surface49}
\end{figure*}

Quantum error correction works by making quantum information highly redundant so that errors affecting a few degrees of freedom become correctable \cite{terhal:rmp}. One can formulate some rough desiderata of a quantum error correcting architecture which aim at minimizing experimental complexity: (1) the architecture has a high noise threshold, providing logical qubits that have a lower logical error probability per logical gate than their physical constituent qubits, (2) it allows for the implementation of a universal~\footnote{A discrete set of gates is universal if the gates in the set can be composed to make any desired unitary gate up to arbitrary precision.} set of logical gates, and (3) it achieves these goals with low spatial (number of physical qubits per logical qubit) and temporal overhead (time duration of logical gate versus physical gate). In addition, (4) it should be possible to process error information sufficiently fast, keeping up with the advancing quantum computation. A last desired property (5) may be that the code is a LDPC (low-density parity check) code: each parity check\footnote{Parity checks operators are operators that define the code space and return a +1 outcome when measured in the absence of errors.}  involves at most $k$ qubits (parity check weight $k$) and each qubit participates in at most $l$ parity checks (qubit degree $l$) where both $l$ and $k$ are small constants. Strongly preferred for solid-state qubits is \etc{a code for which the parity checks act on neighboring qubits in a 2D or 3D array.}



\begin{figure*}[bt]
 \begin{shaded*}
 \raggedright
    \underline{Box \boxnum\label{box:steanecode} $\vert$ \textbf{Steane's Code}} 
\begin{wrapfigure}{r}{0.3\textwidth}
	\includegraphics[width=0.28\textwidth]{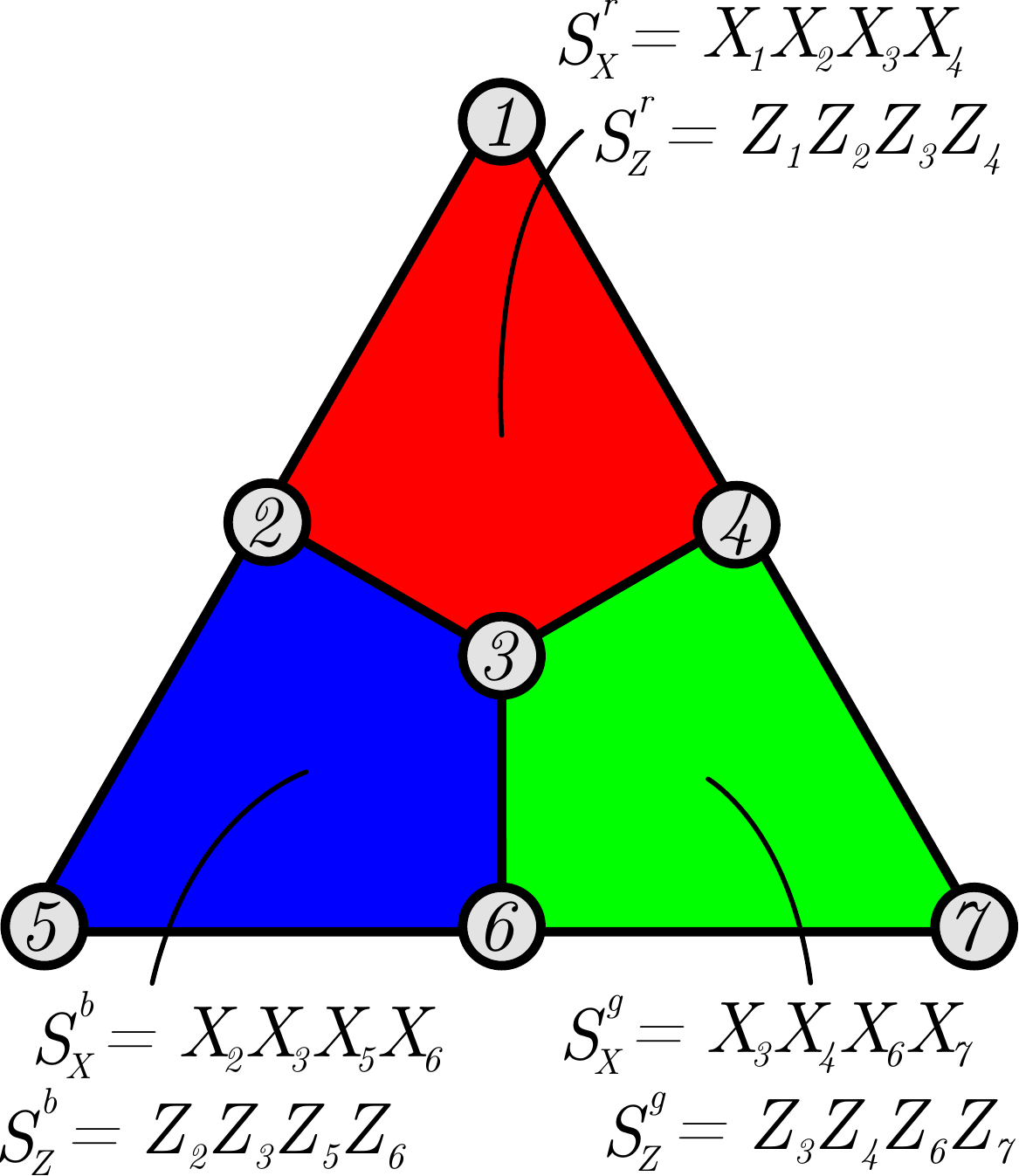}
\end{wrapfigure}
Andrew Steane's 7-qubit code [[7,1,3]] is the smallest example of a 2D color code encoding a single logical qubit $k=1$. The code is defined as the $+1$ eigenspace of the 3 sets of commuting $X$-parity checks $S^r_X, S^g_X, S^b_X$ and $Z$-parity checks $S^r_Z, S^g_Z, S^b_Z$, acting on the 7 qubits located at the vertices. Quantum error correction proceeds by nondestructively measuring these parity checks using ancilla qubits. The $X$-operator of the logical qubit can be chosen as $\Pi_{i=1}^7 X_i$ and the $Z$-operator is $\Pi_{i=1}^7 Z_i$, both of which can be reduced to products of 3 Pauli operators (parity check weight 3) by multiplication with check operators. The code distance is thus 3. The symmetry between $X$ and $Z$-checks ensures that Hadamard \etc{ and $S$} gates can be implemented transversally. Seven CNOT gates between two blocks of Steane code realize the logical CNOT as can be verified from its action on the logical operators and the stabilizer group generated by the parity check operators. It was shown\cite{KLZ:arxiv} that one can implement a fault-tolerant controlled-$S$ gate between two Steane blocks by applying 7 rounds, each with 7 block-wise Controlled-$S$ gates, each round followed by $X$ error correction. This makes for a fault-tolerant universal set of gates.  \etc{The Steane code has been implemented in  ion-trap qubits\cite{nigg:steanecode}}.


\end{shaded*}
\end{figure*}

An important universal gate set is the Clifford+$T$ set. The subset of Hadamard $H$, CNOT and $S={\rm diag}(1,e^{i \pi/2})$ are Clifford gates. A quantum \etc{circuit} comprising only Clifford gates is not universal and confers no quantum computational advantage as it can be classically simulated by the Gottesman-Knill theorem~\cite{thesis:gottesman,AG:stabilizer}. When single-qubit gates come about through resonant driving fields, rotating the qubit vector around its Bloch sphere, a $T={\rm diag}(1,e^{i \pi/4})$ gate is similar in complexity to an $S$ or $H$ gate. For a logical qubit, say,  the one encoded by Steane's 7-qubit code (Box~\ref{box:steanecode}), the logical Hadamard is implemented by applying a Hadamard gate on each of the seven physical qubits. This is advantageous since it takes the same time as an elementary Hadamard gate and the transversal~\footnote{For a single block of code transversal logical gates are realized as a product of single-qubit unitary gates. Transversal logical gates between multiple blocks may use non-product unitary gates provided that these interactions are only between the different blocks.} character of the logical gate ensures that errors do not spread between qubits of the code. The $S$ gate and the CNOT gate are similarly transversal for the Steane code\etc{, but the $T$ gate is not.  Certainly} {\em some} sequence of single and two-qubit gates can be designed to enact a logical $T$ gate. While this is true, the presence of two-qubit gates in such a $T$ gate construction will entirely negate the benefits of using a logical qubit. A sequence of two-qubit gates can spread correctable single-qubit errors to uncorrectable multi-qubit errors, \etc{making} the logical qubit error probability higher than the error probability of a single constituent qubit.

Transversal logical gates are the easiest example of fault-tolerant logical gates, meaning logical gates which do not convert correctable errors into uncorrectable ones.
Transversal gates are optimal in both spatial and temporal overhead. However, it was proved \cite{Chen:earlyEK, EK:nogo} that no non-trivial code allows for the transversal implementation of all gates needed for universality, demonstrating the need for other constructions. 

\etc{A promising architecture uses the surface code,} which was first put forward as a topological quantum memory\cite{dennis+:top}. It has a high noise threshold \etc{$p_c \approx 0.6-1\%$}\cite{RH:topoPrl,fowler+:unisurf,FMMC:review} and requires only a 2D qubit connectivity with qubit degree 4. One logical qubit comprises $d^2$ physical qubits (plus $d^2-1$ ancilla qubits for parity check measurements) for a code distance $d$, see Fig.~\ref{fig:surface49}.  Besides this encoding, there are at least two other ways of defining logical qubits in the surface code. A logical qubit can comprise two holes in an extended surface code sheet \cite{RHG:threshold} or a logical qubit could be represented by two pairs of lattice defects or twists \cite{bombin:twist, HG:dislocation}.  The logical error probability per round of parity check measurements $P_L$ is determined by the distance, i.e.  $P_L \propto (p/p_c )^{d/2}$. Numerical studies\cite{FMMC:review} estimate that, assuming a depolarizing error probability $p<10^{-3}$ per elementary gate, a logical qubit will consist of more than $10^4$ physical qubits in order for $P_L< 10^{-15}$.

\begin{figure*}[thb]
 \includegraphics{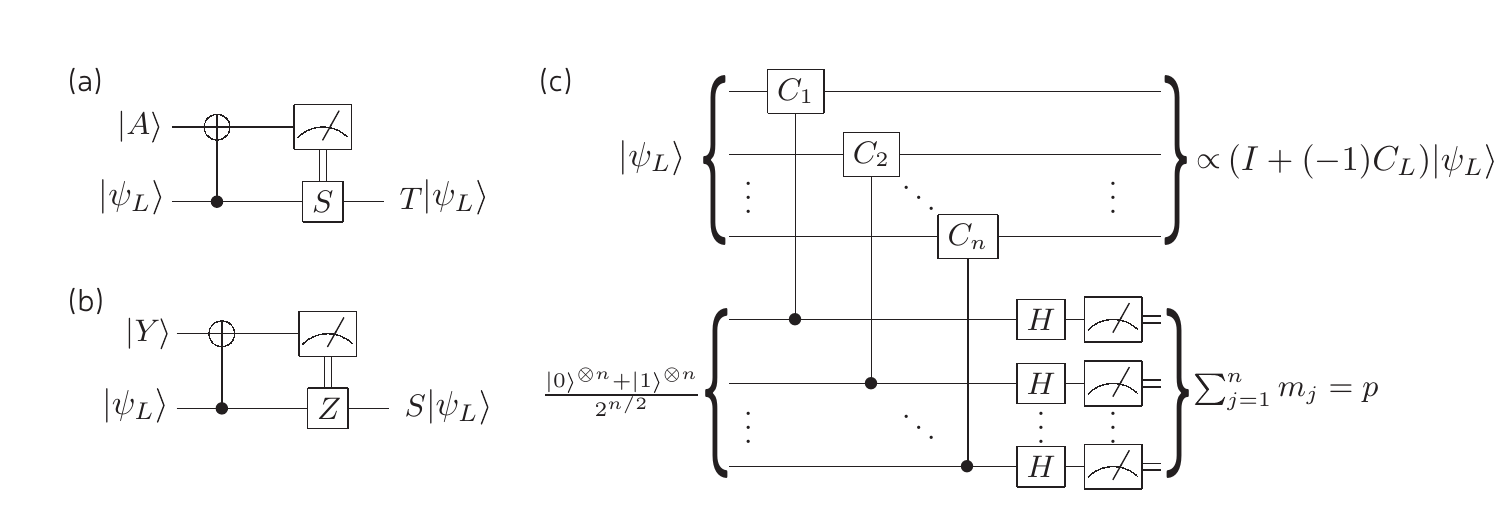}
\caption{(a) Implementation of a $T$ gate via preparing the magic ancilla $\ket{A}=T H\ket{0}$. (b) Implementation of a $S$ gate via preparing the magic ancilla $\ket{Y}=S H\ket{0}$. (c) Fault-tolerant implementation of the projection of a code state onto an eigenstate of a transversal logical Clifford gate $C_L = \prod_{j=1}^nC_j$ assuming that the Clifford gate $C$ has $\pm 1$ eigenvalues. To prepare a $T$ magic state one takes $C = TXT^\dagger \propto SX$, using a transversal logical $S$ gate for the base code. Since a single error can flip the outcome $p$, the circuit has to be repeated and a majority vote over the outcomes taken.}
\label{fig:Ttele}
\end{figure*}

How does the surface code architecture handle the fault-tolerant implementation of gates?  This is partially achieved by code deformation, which is a versatile technique used in fault-tolerantly executing the logical gates:  The code is altered by changing which parity check measurements are done where. A logical CNOT can be obtained with the encoding in Fig.~\ref{fig:surface49} through a deformation technique called lattice code surgery \cite{horsman+:suture}, see Fig.~\ref{fig:surface49}b and \ref{fig:surface49}c.  The encoding of Fig.~\ref{fig:surface49} can also be deformed to a twist defect encoding where the four corners of the lattice correspond to four defects\cite{brown+:poking}. Braiding of the four twist defects through the bulk of the lattice then generates the single-qubit $S$ and Hadamard gate \cite{brown+:poking}. Hence all Clifford gates could be done in situ on 2D surface code sheets.  Another common approach to logical $S$ gates is shown in Fig.~\ref{fig:Ttele}b and uses state injection of a $\ket{Y}\propto \ket{0}+i\ket{1}$ state, which is an inexpensive resource as it may be used many times~\cite{Aliferis07,jones:reuseable}.





\section{Towards Universality}


\subsection{First Ideas}

The necessity of finding effective means to implement a universal set of gates was realized from the earliest beginnings of the field. One useful tool is the simple circuit, shown in Fig.~\ref{fig:Ttele}a, where we replace executing the $T$ gate by preparing the ancilla state $\ket{A}=T H\ket{0}$, called a $T$ magic state. The circuit can be executed at the logical level where one \etc{encodes} qubit and ancilla in a base code.  
Peter Shor provided the first construction\cite{shor:faulttol} for fault-tolerantly preparing a Toffoli magic state. A similar construction\cite{KLZ:res,KLZ:arxiv} was proposed for \etc{fault-tolerantly} preparing a $T$ magic state.  In these constructions one uses the fact that the logical magic state is an eigenstate of a logical Clifford gate which is transversal for the base code. Ancillary cat states are used to fault-tolerantly project onto this eigenstate, see Fig.~\ref{fig:Ttele}c\etc{, but the approach does not scale favourably for topological codes.} 




 \begin{figure*}[t]
 \centering
  	\includegraphics[width=0.9\textwidth]{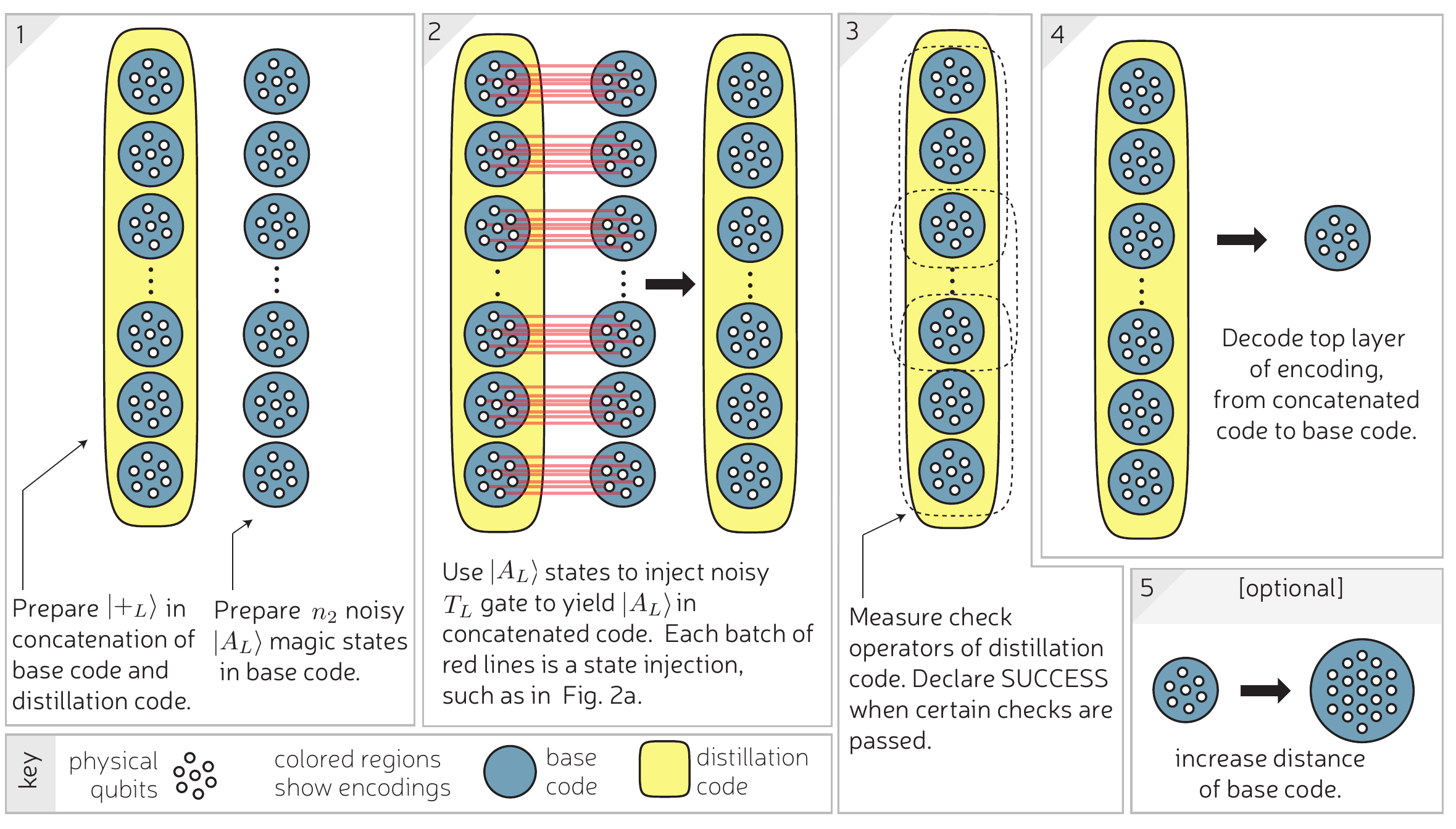}
 \caption{Sketch of magic state distillation using a $[[n_2, 1, d_2]]$ distillation code with a transversal $T$ gate, while Clifford operations are protected by a $[[n_1, 1, d_1]]$ base code. Given fewer than $d_2$ errors in the noisy magic states, they are detected in step 3.  This implies that the logical error probability is suppressed from $\epsilon$ to $O(\epsilon^{d_2})$.  Iterating $r$ times, the error probability reduces to $O(\epsilon^{d_2^r})$.}
 \label{fig:magic}
 \end{figure*}

\subsection{Magic State Distillation}

\begin{figure*}[htb]
\centering
\includegraphics{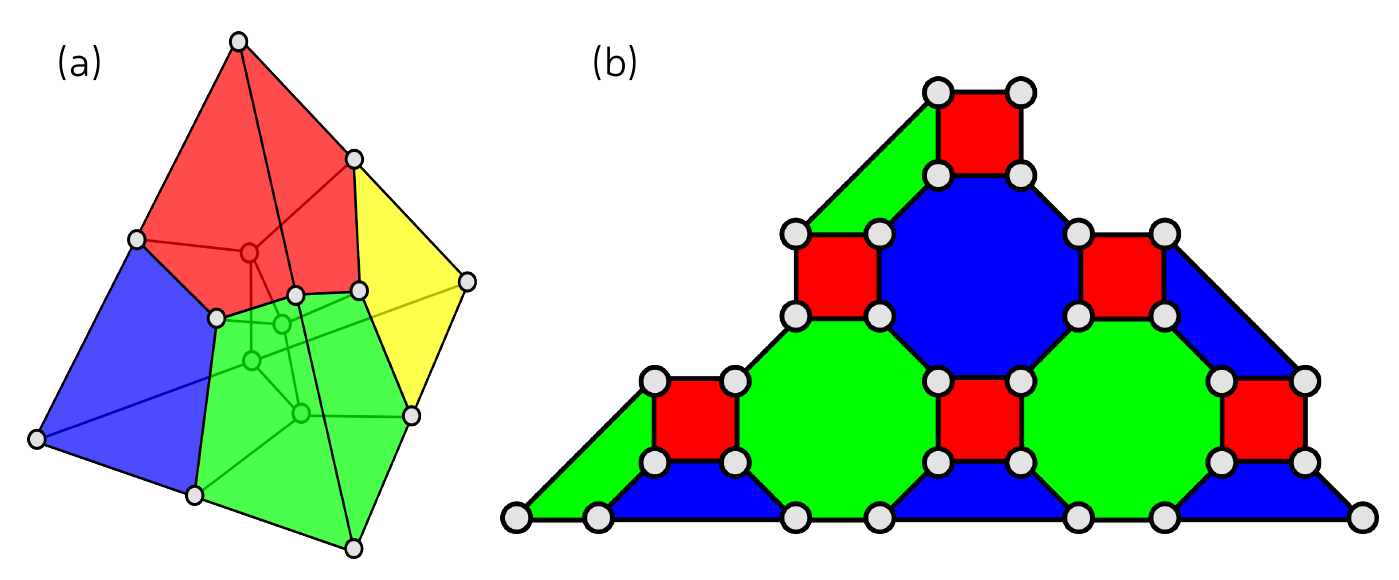}
\caption{(a) Smallest example $[[15,1,3]]$ of a tetrahedral 3D color code with qubits on the vertices. Each colored cell corresponds to a weight-8 $X$-check and each face corresponds to a weight-4 $Z$-check.  A logical $Z$ is any weight-3 $Z$-string along an edge of the entire tetrahedron. The logical $X$ is any weight-7 $X$-face of the entire tetrahedron. A logical $T^\dagger$ is implemented by applying a $T$ gate on every qubit.  The online Supplementary Information\cite{SuppMat::colorcode} has a movie of a larger 3D color code. (b) A 2D triangular [[31,1,7]] code, generalizing Steane's code, based on a 4.8.8. lattice. The qubits are associated with vertices and each colored face correspond\etc{s} to both an $X$ and a $Z$-check. A logical $X$ or $Z$ is a $X$-string (resp. $Z$-string) running along any of the edges of the entire triangle.}
\label{fig:colorcodes}
\end{figure*}

The mindset of Magic State Distillation (MSD) is to accept a preparation procedure providing noisy magic states, and then proceed by filtering many noisy magic states into fewer, yet better quality states.
Efficient protocols for magic state distillation are designed, almost exclusively, using an error correction code with a transversal non-Clifford gate, typically the $T$ gate. We call this code the distillation code. \etc{The $[[15,1,3]]$ quantum Reed-Muller code was shown\cite{KLZ:arxiv} to have a transversal $T$, and using this code for magic state distillation was proposed by Bravyi and Kitaev\cite{BK:magicdistill}.}  The $[[15,1,3]]$ code is now also recognized as the smallest member of a 3D color code family, see Fig.~\ref{fig:colorcodes}. All MSD protocols work at the logical level of an underlying base code and so assume reliable Clifford operations. In Fig.~\ref{fig:magic} we outline one variant of MSD.  

One figure of merit is the number of noisy magic states consumed per single $T$ gate.
Advances in distillation codes have improved the asymptotic efficiency by this metric (see Box~\ref{box:MSD}).
However, the number of physical qubits involved (space cost) and protocol duration (time cost) are more realistic metrics, although they depend on the choice of base code. When optimizing full space-time costs, an important trick is to increase the base code distance with successive distillation rounds (see step 5 of Fig.~\ref{fig:magic}).
Using this trick in conjunction with the surface code, resource overheads become dominated by the surface code cost in the final distillation round~\cite{RHG:threshold,FMMC:review,fowler:blockMSD,Ogorman:realisticMSD}.
This results in a space-time cost for the $T$ gate \etc{that} is only a constant multiple of a surface code overhead, namely a $O(d^2)$ spatial cost and a $O(d)$ temporal cost. More precisely, the space-time cost of a $T$ gate realized in a distance-$d$ surface code is $C_T d^3$ with $C_T\approx 160-310$ when employing Bravyi-Haah codes~\cite{FMMC:review,fowler:blockMSD,Ogorman:realisticMSD}. For Clifford gates the overhead per logical gate is also $O(d^3)$ but the constant prefactor is of order of unity.  Using even higher yield MSD protocols (see Box~\ref{box:MSD}) may reduce the $C_T$ factor further, with the Bravyi-Haah codes already shown~\cite{fowler:blockMSD} to have three times lower space-time costs than [[15,1,3]].



Obtaining a fault-tolerant logical $T$ gate is only a partial goal as Clifford+$T$ gates are then used to synthesize other logical gates needed in quantum algorithms~\cite{KSV:computation,RS14,bocharov:PQF,amy:TPAR}.
A more efficient solution can be to directly distill magic states providing the most frequently required logic gates.
Magic state distillation has been shown for smaller angle $Z$ rotations~\cite{landahl13,duclos15,campbell16}, the Toffoli gate~\cite{eastin13,jones13b}, and a general class of multi-qubit circuits~\cite{campbell:unified2}.  

Given that magic state distillation takes up space and time, it requires an allocation of resources and communication infrastructure (in the form of logical roads) to these resources inside the 2D surface code architecture. Clifford gates in such \etc{an} architecture could be done in situ on a `Clifford substrate' of 2D surface code sheets.
Throughout this 2D array of sheets, areas are reserved for magic state distillation factories for non-Clifford gates. The optimal spatial density of these factories depends on the typical quantum algorithmic use (frequency, parallelism) of non-Clifford gates. Clearly, any design of such a quantum computer will require a huge effort in integrated quantum circuit design and optimization, a quantum analog to VLSI design. This effort has barely gotten underway\cite{PDF:circuit}.

\begin{figure*}[bht]
 \vspace{-10pt}
\raggedright
\begin{shaded*}
\noindent
\underline{Box \boxnum\label{box:MSD} $\vert$ \textbf{High yield MSD}} 

\begin{wrapfigure}{t}{0.5\textwidth}
\vspace{-20pt}
 \begin{tabular}{lcc}
\rowcolor{headerrow}
 	Protocol	       			          & $[[n,k,d]]$ & $\lim_{n\rightarrow \infty}  \gamma$  \\
 Reed-Muller  			 	              & $[[15,1,3]]$ &   2.464 \\
 Meier-Eastin-Knill~\cite{Meier13}        & $[[10,2,2]]$ &   2.322      \\
 Bravyi-Haah~\cite{BH:magic}              & $[[3k+8,k,2]]$ & 1.585 \\
 Jones $2^{\mathrm{nd}}$ level~\cite{jones:dist}	  & $[[5k + O(1), k ,4]]$  & 1.160 \\
 Jones $r^{\mathrm{th}}$ level~\cite{jones:dist}	  & $[[ (2^r+1)k + O(1), k, 2^r ]]$  & 1 
\end{tabular}
\end{wrapfigure}
\vspace{5 pt}
The yield is the number of distilled magic states, on average, per input noisy magic state. Using an $[[n_2,k_2,d_2]]$ distillation code for enough rounds to achieve some target $\epsilon_\mathrm{out}$ has an asymptotic yield $1/O( \log(\epsilon_\mathrm{out}^{-1})^{\gamma} )$ where $\gamma = \log_{d_2}(n_2/k_2)$.  Therefore, lower $\gamma$ values indicates more efficient distillation protocols.  More sophisticated protocols (see table) can reduce the scaling factor $\gamma$, with $\gamma=1$ conjectured~\cite{BH:magic} to be optimal.
\end{shaded*}
\end{figure*}


\begin{table*}[htb]
\begin{center}
\begin{tabular}{ccccccc}

\rowcolor{headerrow}
 &   & Parity Check  & Threshold    & Threshold & &  \\
\rowcolor{headerrow}
Code &  Qubit Degree &  Weight & (Phen. Model)   &  (\etc{Circuit} Model)   &  Single-Shot & Logic \\
 2D Surface&  4 & 4 & $2.9\%$\cite{WHP:threshold} & $0.6\%$\cite{RH:topoPrl}-$1\%$\cite{fowler+:unisurf,FMMC:review} & No & Clifford \\  
 2D 6.6.6 Color & 6 & 6 & $2.8\%$ \cite{beverland:perscom,beverlandThesis} & $0.3 \%$\cite{beverland:perscom,beverlandThesis} & No & Clifford \\
 3D Gauge Color & 12 & 6 &  $0.31\%$ \cite{brown:singleshot}  & Unknown & Yes & Clifford \\
 3D Color& 10 & 24 & Unknown & Unknown & No & Transversal $T$ \\
  4D Surface& 8 & 6 &  $1.59\%$ \cite{BDMT:local} & Unknown & Yes & $^\dagger$  
    \end{tabular}
\end{center}
\caption{Parity check weight is given for the bulk of the code lattice. \etc{The threshold depends on noise model: The phenomenological model assigns probability $p$ equally to $X$ \& $Z$ errors and an error in the parity check measurements; The circuit model applies depolarizing noise with probability $p$ to every elementary component in the circuit implementing the parity checks. The phenomenological threshold is always higher than circuit model threshold, especially} for codes with high parity check weight.  $^\dagger$It has been shown\cite{KYP:unfolding} that one can perform a \etc{fault-tolerant} non-Clifford 4-qubit-controlled-Z using a \etc{constant depth circuit}.}
\label{table:codes}
\end{table*}

\subsection{Color Codes}

The [[7,1,3]] and [[15,1,3]] codes are the smallest members of a family of 2D and 3D, respectively, color codes\cite{BM:colorcodes,BM:transT,bombin:homological}. Examples are given in Fig.~\ref{fig:colorcodes}.
These color codes retain the transversality properties of their respective smallest instances. \etc{Therefore,} the 2D color codes have transversal Clifford gates~\cite{katzgraber:fullClifford,KB:simple,bombin:gauge}. The 3D color codes can have a transversal non-Clifford \etc{gate}~\cite{BM:transT,KYP:unfolding}, such as a $T$ gate in the tetrahedral 3D color code. Lattice code surgery \cite{landahl14} can again be used to locally perform CNOT gates.
Color codes can also be extended to higher dimensions (see Box~\ref{box:CC}), with transversality properties related to the dimensionality. 

The 3D color code does not have the symmetry between $X$- and $Z$-checks, and \etc{so} lacks a transversal Hadamard gate. Ideas to get around this caused a surge of interest in 3D color codes. One idea is that of switching between different (3D color) codes via the important concept of gauge fixing~\cite{paetznick:gauge,bombin:gauge}. Using gauge fixing it is possible to use the transversal $T$ gate of the 3D color code while only doing error correction and the Hadamard gate with the 3D gauge color code\cite{bombin:singleshot}. The advantage of using the 3D gauge color code over the 3D color code, see Table \ref{table:codes}, are the lower parity check weights and the feature of single-shot error correction of the 3D gauge color code (see Sec.~\ref{Sec:singleshot}). A CNOT gate realized via lattice code surgery allows for the injection of the $T$ gate from a 3D gauge color code into a 2D color code~\cite{VT:color_arch,bombin:dimensionalJumps}. \etc{One can imagine} a 2D color code architecture augmented with 3D gauge color code $T$-stations where logical qubits can undergo a $T$ gate, similar as a 2D surface code with 2D non-Clifford processing occurring at dedicated locations.

We summarize some of the known thresholds and properties of codes in Table \ref{table:codes}. Note that the 3D color and gauge color codes have a cubic spatial overhead $O(d^3)$ for a given distance $d$ while this overhead is $O(d^2)$ for 2D codes.  The complexity of decoding 3D color and 3D and 4D surface codes poses new challenges and is not fully understood while good algorithms exist for surface code decoding.
The best thresholds for 2D color codes are lower than those of the surface code, possibly due to the fact that parity checks have higher weight.
However, 2D color code decoding is also more computationally complex than surface code decoding \cite{delfosse:colordecoding}.
The best threshold numbers for circuit-based noise in which each gate undergoes depolarizing noise with probability $p$ \etc{are $0.3\%$\cite{beverland:perscom,beverlandThesis} for} a triangular color code and $0.41\%$ for a half-color or $[[4,2,2]]$-concatenated toric code \cite{CT:422} (compare with $0.6-1\%$ for the surface code).


\begin{figure*}[hbt]
 \begin{shaded*}
 \raggedright
    \underline{Box \boxnum\label{box:CC} $\vert$ \textbf{Arbitrary dimension color codes}} \\ \vspace{5 pt}
The general construction $D$-dimensional gauge or stabilizer color codes is based on a $D$-dimensional simplicial complex whose vertices are $D+1$-colorable (adjacent vertices have different colors).  One associates qubits with $D$-simplices, $X$- and $Z$-parity checks with respectively $x$- and $z$-simplices, obeying $x+z \leq D-2$. This inequality and the colorability property enforce the commutation of the checks, leading to a code family ${\rm ColorCode}_D(x,z)$ \cite{bombin:gauge,KB:simple}. A common boundary configuration for such codes is obtained by tiling the inside of a big $D$-simplex, respecting colorability but omitting elements of the bigger simplex as possible checks (see Suppl. Inf.\cite{SuppMat::colorcode} for the construction of a 3D code). Such color codes encode one logical qubit and we can call them simplicial color codes.\\
\indent
In 2D qubits are on triangles and the only choice is $x=z=0$. This implies that a $X$- and a $Z$-parity check is associated with each vertex, \etc{leading} directly a transversal $H$ gate. The simplicial (or triangular) version also has a transversal $S$ gate. The common representation of such codes, Fig.~\ref{fig:colorcodes}, is obtained by going to the dual lattice where qubits are associated with vertices and checks are associated with faces of the lattice. \\
\indent
For dimensions higher than 2D, it is possible to choose $x+z<D-2$. In that case, the code space contains gauge qubits whose $X$- and $Z$-gauge checks are given respectively by $(D-2-z)$- and $(D-2-x)$-simplices. Remarkably, the gauge checks redundantly represent the stabilizer check information giving rise to single-shot error correction, discussed in \ref{Sec:singleshot}. 
\indent
In 3D qubits are on tetrahedra and one has three choices for $(x,z)$.  Choosing $(x,z)=(0,0)$ gives a 3D gauge color code \cite{bombin:gauge}. This code has a \etc{$X$- and $Z$-stabilizer} check for each vertex leading to a transversal Hadamard and gauge checks associated with edges. Another option is the 3D color family with $(x,z)=(0,1)$ in which the $X$-checks are associated with vertices and the $Z$-checks with edges. The simplicial (or tetrahedral) version of this 3D color code has a transversal $T$ \cite{BM:colorcodes} and the smallest example [[15,1,3]] is shown in Fig.~\ref{fig:colorcodes}a. The third option $(x,z)=(1,0)$ is trivially (Clifford) equivalent to the $(x,z)=(0,1)$ codes. 
\end{shaded*}
\end{figure*}


\subsection{Alternative Code Constructions}
\label{Sec::AltCode}



An alternative to topological error correction is concatenated coding in which the physical qubits in a code block are repeatedly replaced by logical qubits.  Extensive work \cite{CDT:study} has been performed on comparing the overheads and noise thresholds of various schemes. For a (concatenated) [[23,1,7]] Golay code (with a transversal Clifford set) it has been shown that the asymptotic noise threshold is at least $0.13\%$ \cite{PR:golay} (compare with the numerical value $0.6-1.0\%$ of the asymptotic surface code). Any concatenated scheme with easy Clifford gates could be combined with magic state distillation. The performance comparison with the surface code would largely rely on how much spatial overhead one pays for a logical Clifford gate.

Another concatenation idea is to combine the transversality of different gates in two different codes \cite{oconnor_laflamme} and get rid of magic state distillation. For example, one can choose [[7,1,3]] as a top code, i.e. replacing each physical qubit by 7, and then take [[15,1,3]] as bottom code, replacing each of the 7 qubits again by a block of 15. The resulting code is $[[105,1, 9]]$. Due to the non-transversality of the Hadamard at the bottom level and the non-transversality of the $T$ gate at the top level, the total logical error probability of these gates will suffer, but single-qubit errors can still be corrected in this construction. The asymptotic noise threshold of this construction was lower-bounded by $0.28\%$ \cite{chamberland:analysis}.


In Box~\ref{box:steanecode} it was stated that the Steane code has a {\em pieceable fault-tolerant} \cite{YTC:piece} Controlled-$S$ gate. This means that we can break down the execution of the gate in rounds or pieces, each round containing $X$ error-correction to maintain fault-tolerance, but holding off on $Z$ error correction until the entire gate is done. This idea does not easily scale to topological codes, but it could be analyzed for concatenated codes. It obviates the need for magic state distillation, but trades this, mostly likely, with a poorer asymptotic noise threshold.


 
Any scheme based on the concatenation of small codes can be converted to a coding scheme \etc{that} is local in 1D or 2D at the cost of some additional overhead for gates which move qubits. If a large 3D color code is implemented in pure 2D hardware, it requires non-planar connections whose length grows with the size of the color code. Recent work \cite{bravyi:doubled, JBH:2D, jochym:stacked} has shown how to systematically construct codes that have a transversal $T$ gate and convert these codes to so-called doubled color or 2D gauge color codes. However, by making all connections local on a 2D lattice the resulting 2D codes are non-topological. This means that the code performance is maximal for a certain code size and declines for larger code sizes. The performance of doubled color or 2D gauge color codes in producing a low-noise logical $T$ ancilla (which can then be transferred to the 2D Clifford substrate) has not yet been compared with MSD or the usage of 3D $T$ stations.

 


\subsection{Comparison of Resource Overheads}
   
The combination of MSD with the surface code is currently considered a competitive scheme since it combines a high-noise threshold, a 2D architecture and a $T$ gate \etc{that is a few hundred times} as costly as Clifford gates in terms of its space-time overhead.  Replacing MSD by 3D $T$-gate stations and/or the surface code substrate by a color code substrate is an alternative whose appeal depends on the physical error probability versus the 2D and 3D color code thresholds. The use of the 3D gauge color code for a $T$ gate requires $O(d^3)$ qubits, but single-shot error correction makes the space-time cost again $O(d^3)$\cite{bombin:singleshot}.



An analysis\cite{chamberland:analysis} of the concatenation scheme discussed in Sec.~\ref{Sec::AltCode} shows that the spatial overhead is still not favorable as compared to \etc{using surface codes with MSD}. This analysis includes the consideration of a smaller, more efficient, 49-qubit code\cite{nikahd:nonuniformCC}. To get to the target error probability $P_L < 10^{-15}$ starting with a physical error probability of $O(10^{-5})$ it is estimated that the concatenated scheme uses at least $10^7$ physical qubit for a logical qubit (versus $10^4$ for \etc{surface codes with MSD}).



\section{The Blessing of Dimensionality?}

\subsection{Transversality and Dimensionality}

A deep connection between transversality and dimensionality of a topological stabilizer code was proved by Bravyi and Koenig \cite{Bk:uni}. 
Their theorem says that for $D$-dimensional topological stabilizer codes, the only logical gates that one can implement via a transversal or constant depth circuit are in the so-called $m^{\mathrm{th}}$ level of the Clifford hierarchy ${\cal C}_m$.
Here ${\cal C}_1$ is the group of $n$-qubit Pauli operators, ${\cal C}_2$ is the Clifford group, ${\cal C}_3$ contains gates such as $T$ and Toffoli and the $m^{\rm th}$ level includes a small rotation gate such as ${\rm diag}(1,\exp(\frac{2 \pi i}{2^{m}}))$.
While the gate set Clifford+$T$ is universal, it can be shown that using gates in higher levels of the Clifford hierarchy reduces the time overhead associated with gate synthesis.
The gates that one can realize with $D$-dimensional color and surface codes saturate the Bravyi-Koenig theorem \cite{bombin:gauge,KB:simple,KYP:unfolding}. 
This theorem does not prove that there are no good 2D alternatives to magic state distillation.
It does suggest that any alternative might come at the price of a lower threshold for universal quantum logic as it requires a non-trivial fault-tolerant gate construction.



\subsection{Tradeoff Bounds}

For the design of a storage medium, a quantum hard-drive, one can drop the universal gate set desideratum (2) as long as quantum information can be read and written to storage.
Ideally, the storage code is a $[[n,k,d]]$ code with high rate $k/n$ and high distance $d$ scaling as some function of $n$.
It was shown~\cite{BPT:tradeoff} that 2D Euclidean topological stabilizer codes are constrained by $k d^2 \leq c n$ for some constant $c$: the surface code clearly saturates this bound.
The adjective Euclidean means that the qubits can be placed on a 2D regular grid with each qubit connecting to O(1) neighbors. \etc{Consequently,} the rate of these codes vanishes with increasing distance, leading to a substantial overhead as a storage code.
Hyperbolic surface codes \cite{BT:hyperbolic} are only bound by $k d^2 \leq c (\log k)^2 n$ \cite{delfosse:tradeoffs}.
There is a simple hyperbolic surface code in which qubits are placed on the edges of square tiles and five tiles meet at a vertex.
Such $\{5,4\}$-hyperbolic surface codes have an asymptotic rate $\frac{k}{n}=\frac{1}{10}$ and logarithmically growing distance\cite{BT:hyperbolic}.






\subsection{Single-shot Error Correction}
\label{Sec:singleshot}
2D topological codes have an intrinsic temporal overhead in executing code deformation in a fault-tolerant manner, making gates that rely on this technique take $O(d)$ time.
The reason is that in code deformation new parity check measurements are repeated $O(d)$ times in order for this record to be sufficiently reliable.
The absence of redundancy in the parity check measurement record is an immediate consequence of the lack of self-correction of 2D topological stabilizer codes\cite{BT:mem}.
A 4D hypercubic lattice version of the surface code \cite{dennis+:top} allows for {\em single-shot} error correction instead.
Due to redundancy in the parity check data in this code, it is possible to repair the data for measurement errors after a single round of measurement. Codes which have such single-shot error correction then have potentially higher noise thresholds and faster logical gates. 
Interestingly, Bombin showed that single-shot error correction is possible for 3D gauge color codes \cite{bombin:singleshot}. For such code the value of the stabilizer parity checks is acquired through measuring non-commuting lower-weight gauge qubit checks whose products determine the stabilizer parity checks.
Curiously, in the 3D gauge color code, the gauge qubit checks hold redundant \etc{information,} which allows one to construct a robust record for the stabilizer parity checks in O(1) time.
The result shows the power of using subsystem codes with gauge degrees of freedom since we do not expect to have single-shot error correction for 3D stabilizer codes.

\section{Outlook}

\etc{We have discussed several ideas for adding universal computing power to a quantum device. Presently, using surface codes with magic state distillation is the most practical solution.  We have seen that there is a wealth of fascinating alternatives, but so far they have yet to demonstrate a comparably high threshold or significant improvements in resource scaling. A interesting direction is to move away from the constraints of low dimensional topological codes.}

More general LDPC codes could be considered, for example the 4D surface code in Table \ref{table:codes}. Homological quantum (LDPC) codes can in principle be constructed from tilings of any $D$-dimensional manifold. Generalizations of classical LDPC codes based on expander graphs to quantum codes are also known to exist \cite{TZ:codes, FH:hypergraphs}.

Such approaches require hardware that supports long-range connectivity. Fortunately, various \etc{long-range} experimental platforms such as ion-trap qubits or nuclear spins coupled to NV centers in diamond, do not necessarily conform to the paradigm of a 2D `sea' of qubits. One may expect such architectures to work with modular components with photonic interconnects \cite{monroe+:arch, NFB:photons}\etc{, which would allow for more flexible and long-range connectivity.}



The advantage of using a higher-dimensional LDPC code or more general quantum LDPC code for computation or storage in a concrete coding architecture remains to be fully explored, in particular efficient decoding software needs to be developed. Independent of whether these codes can be used in a coding architecture, we expect that the study and development of quantum LDPC codes will lead to new insights into robust macroscopic quantum information processing.


\section{Acknowledgements}
BMT and CV acknowledge support through the EU via the ERC GRANT EQEC and ETC is supported by the EPSRC (EP/M024261/1). 


%

\end{document}